\documentclass[12pt]{article}
\usepackage{amssymb,graphicx}
\usepackage{epstopdf}
\usepackage{amsmath,amsfonts}
\usepackage{epsfig}

\newcommand{\be}{\begin{equation}}
\newcommand{\ee}{\end{equation}}
\newcommand{\bea}{\begin{eqnarray}}
\newcommand{\eea}{\end{eqnarray}}
\newcommand{\bg}{\begin{gather}}
\newcommand{\eg}{\end{gather}}
\newcommand{\bseq}{\begin{subequations}}
\newcommand{\eseq}{\end{subequations}}

\renewcommand{\exp}{\mathop{\rm exp}\nolimits}
\renewcommand{\ln}{\mathop{\rm ln}\nolimits}

\begin{document}

\begin{center}

 {\LARGE \bf Galileon bounce \\ after ekpyrotic contraction } \\
\vspace{20pt}
%\medskip
M.~Osipov$^{a}$,~V.~Rubakov$^{a,b}$\\
\vspace{15pt}

$^a$\textit{
Institute for Nuclear Research of
         the Russian Academy of Sciences,\\  60th October Anniversary
  Prospect, 7a, 117312 Moscow, Russia}\\
\vspace{5pt}
$^b$\textit{
Department of Particle Physics and Cosmology, Faculty of Physics,\\ Moscow State University,
Vorobjevy Gory, 119991 Moscow, Russia}
    \end{center}
    \vspace{5pt}

\begin{abstract}

We consider a simple cosmological model that includes a long ekpyrotic contraction stage and smooth bounce after it. Ekpyrotic behavior is due to a scalar field with a negative exponential potential, whereas the Galileon field produces bounce. We give an analytical picture of how the bounce occurs within the weak gravity regime, and then perform numerical analysis to extend our results to a non-perturbative regime.

\end{abstract}

\section{Introduction}

The Universe possesses several features that are difficult to explain within the hot Big Bang theory. An elegant solution to these problems is inflation~\cite{Starobinsky, Guth, Linde, Albrecht}. However, alongside with the inflationary theory there exist a number of cosmological scenarios which can also help to deal with the hot Big Bang theory problems. One set of ideas has to do with ekpyrotic~\cite{Khoury2001, Khoury2002-2, Khoury2004, Creminelli2006, Buchbinder2007-1, Buchbinder2007-2}, bouncing~\cite{Khoury2002-1, Tolley, Creminelli.Bounce, Brandenberger} and cycling~\cite{Steinhardt, Erickson2007, Gasperini-Veneziano, Erickson2004} cosmologies: long enough contraction epoch preceding the hot Big Bang stage could be an alternative to inflation.

One way to consistently obtain the contracting pre-Big Bang stage is to introduce a scalar field with a negative exponential potential: $V_{\phi} \propto -\exp(-c\phi)$. This kind of potential leads to the stiff equation of state, $p/ \rho > 1$. Stiff matter avoids the "chaotic anisotropy" effect~\cite{BKL}; otherwise one would encounter anisotropic collapse of the Universe which is hard to convert into smooth expansion. On the contrary, contracting stage with stiff matter is consistent with isotropic, spatially flat and homogenious Universe~\cite{Erickson2004}.

For the bounce to occur one has to ensure that the initially negative Hubble parameter crosses zero. Its derivative is given by the Raychaudhuri equation:
\be
\dot H = -4\pi G (p + \rho),
\label{Raichaudkhuri}
\ee
so that the bounce requires
$$
p + \rho < 0.
$$
In other words, the bounce needs the dominating matter that violates the Null Energy Condition. The problem is that most types of matter do not behave in this way, while N.E.C.-violating theories are often plagued by istabilities. One known exception is Ghost Condensate, whose instabilities at the bounce stage may be mild~\cite{Buchbinder2007-1, Creminelli.Bounce} and, indeed, a model with initially stiff equation of state and bounce due to Ghost Condensate has been recently constructed~\cite{Brandenberger}. Another way is to introduce the Galileon field~\cite{Nicolis.Galileon, Creminelli.Genesis, Creminelli.Sublum}, or its close relative, a field with Kinetic Gravity Braiding~\cite{KGB-1, KGB-2}. These are scalar fields with special forms of the Lagrangian that lead to N.E.C.-violation but allow to avoid instabilities. Bouncing cosmological models with these fields have been proposed in Refs.~\cite{kitaicy, Vikman}.

Our purpose is to construct a cosmological model which describes the Universe that starts from ekpyrotic contraction stage and proceeds to expansion via smooth bounce produced by the Galileon. In particular, we make sure that the cosmological constant is zero at late times. We find that ekpyrotic and bouncing behavior is quite generic, as it occurs in a wide range of the parameters and initial conditions.

The paper is organized as follows. We start Section~\ref{Bounce} with the description of a simplified model, then we study the system behavior at early times (Section~\ref{scalar_bckgr}) and consider the Galileon that evolves in the background of the scalar field. In Section~\ref{slow_bounce} we introduce the perturbative approach and observe the bounce analytically, whereas in Section~\ref{numerics} we present numerical analysis. In Section~\ref{Exit} we modify the model to ensure that the cosmological constant vanishes at late times and show that the late-time inflationary behavior is avoided in a wide region of the parameter space.

\section{Bounce in a simplified model}
\label{Bounce}

\subsection{The Model}
\label{Model}

We begin with a simple model with ekpyrosis and Galileon bounce; the model is modified and made more realistic in Section~\ref{Exit}. We introduce two scalar fields, the Galileon $\pi$ and the second scalar field $\phi$, both minimally coupled to gravity (mostly negative signature):
$$
S=S_{EH} + S_\phi + S_{\pi} \; ,
$$
where
$$
S_{EH}  = -\frac{M_p^2}{2}\int~d^4x~R\sqrt{-g} \; ,
$$
$$
S_\phi  = \int~d^4x~\sqrt{-g} \left( \frac{1}{2} g^{\mu \nu} \partial_\mu \phi \partial_\nu \phi
- V(\phi) \right) \; ,
$$
\be
S_\pi =  \int~d^4x~\sqrt{-g} \left[ - f^2 \mbox{e}^{2\pi} (\partial \pi)^2
+ \frac{f^3}{\Lambda^3} (\partial \pi)^2 \Box \pi
+ \frac{f^3}{2\Lambda^3} (\partial \pi)^4 \right].
\label{S_pi}
\ee
Here $(\partial \pi)^{2n} = (g^{\mu\nu}\partial_\mu \pi \partial_\nu \pi)^{n}$, $\Box = \nabla^2$. This particular version of the Galileon generically has a problem with superluminal propagation. However, this property is absent near the relevant backgrounds in a slightly modified model~\cite{Creminelli.Sublum}, in which the coefficient of the last term in~\eqref{S_pi} is $(1 + \alpha) f^3 / 2\Lambda^3$ instead of $f^3 / 2\Lambda^3$, where $0 < \alpha < 3$. To simplify formulas below, we set $\alpha = 0$, but the analysis goes through for subluminal $\alpha$. 

We study spatially homogeneous fields in FLRW background
\[
ds^2 = dt^2 - a^2 (t) d{\bf x}^2 \; . 
\]
Since we want to start from contracting Universe, we take the potential of $\phi$ in the following form:
\[
V(\phi) = -V_{0} e^{-c\phi} \; , 
\]
where $V_{0}$ and $c$ are positive parameters. 
Energy densities and pressures of the scalar field and the Galileon are
\be
\rho_{\phi} = \frac{\dot\phi^2}{2} + V(\phi) \; ,
\label{rho_phi}
\nonumber 
\ee

\be
p_{\phi} = \frac{\dot\phi^2}{2} - V(\phi) \; ,
\label{p_phi}
\nonumber 
\ee

\be
\rho_\pi = -f^2 e^{2\pi}\dot\pi^2 + \frac{f^2}{H_*^2}(\dot\pi^4 + 4H\dot\pi^3) \; ,
\label{rho_pi}
\nonumber 
\ee

\be
p_\pi = -f^2 e^{2\pi}\dot\pi^2 + \frac{f^2}{3 H_*^2}(\dot\pi^4 - 4\dot\pi^2\ddot\pi) \; .
\label{p_pi}
\nonumber 
\ee
Note that the potential $V(\phi)$ is unbounded from below. We improve on this point in Section~\ref{Exit}.

\subsection{The Galileon in the background of scalar field}
\label{scalar_bckgr}

If the Galileon field starts at $\pi \to - \infty$ as $t \to -\infty$, while the field $\phi$ evolves from $\phi = \infty$, the Galileon energy density is subdominant, and the cosmological contraction is governed by the scalar field. In this section we study ekpyrotic contracion of the Universe and the Galileon field in the contracting background. We are not able to obtain an exact solution for the Galileon; however, a combination of analytical and numerical analyses helps us to understand its behavior before the bounce. We will use these results in the next section as the initial conditions for the "slow bounce" scenario.

If one switches off the Galileon, it is straightforward to obtain the background solution for the scalar field and scale factor:
\be
\phi(t) = -\frac{1}{c}\ln{ \left[ \left(  1 - \frac{6}{M_p^2 c^2}  \right)  \frac{2}{c^2 V_0 t^2} \right] },
\label{scalar_solution_curved}
\ee
\[
a(t) \propto (-t)^{\frac{2}{c^2 M_p^2}},
\]
\[
H(t) = \frac{2}{c^2 M_p^2 t} \; . 
\]
To study the Galileon behavior, we write down its field equation:
\be
e^{2\pi}(\ddot{\pi} + \dot{\pi}^2) - \frac{2}{H_*^{2}} \dot{\pi}^2 \ddot{\pi} + 3 e^{2\pi} H \dot{\pi} - \frac{1}{H_*^2} \left[ 4H\dot{\pi}\ddot{\pi} + 2 \dot{\pi}^2 (3H^2 + \dot{H}) + 2H\dot{\pi}^3  \right ] = 0.
\label{G_EOM}
\ee
Here
\[
H_*^2 = \frac{2 \Lambda^3}{3f} \; . 
\]
In the case of Minkowski background, the stable solution is~\cite{Creminelli.Genesis} 
\be
e^{\pi} = \frac{1}{H_* (t_* - t)}.
\nonumber
\label{pi_Minkowski_1}
\ee
It is parametrized by a parameter $t_*$. In the case of evolving background with 
\[
H = \frac{P}{t},~~~P = \frac{2}{c^2 M_p^2} 
\]
it appears natural to search for the solution in the form 
\be
e^{\pi} = -\frac{\alpha}{H_* t}
\label{Unstable_Solution}
\ee
with some constant $\alpha$. One finds that eq.~\eqref{Unstable_Solution} is indeed a solution to eq.~\eqref{G_EOM} with 
\[
\alpha = \sqrt{\frac{6P^2 - 8P +2}{2 - 3P}}. 
\]
However, this solution is a repulsor. To see this, we consider small perturbation $\pi \to \pi + \delta\pi$. The linearized equation for $\delta\pi$ has the following form:
\[
t^2 \delta\ddot\pi (\alpha^2 - 2 + 4P) + t\delta\dot\pi ( 4 - 2\alpha^2  + 3\alpha^2 P  + 12P^2 - 2P) + \delta\pi (4\alpha^2 - 6\alpha^2 P) = 0  \; . 
\]
To simplify calculations we assume that $c \gg M_p^{-1}$, work to the linear order in the small parameter $P = 2(c M_p)^{-2}$ and get the following solution:
\[
\delta\pi = \frac{C_1}{t^{1-\frac{19}{5}\cdot P}} +  C_2 \cdot t^{4+\frac{34}{5}\cdot P}.
\]
In the case of Minkowski background ($M_p \to \infty,~~P \to 0$), the growing part of perturbation is proportional to $1/t$. It can therefore be viewed as the result of timeshift: if one makes a shift $t \to t + \epsilon$, the Galileon transforms into $\pi \to \pi + \dot\pi\epsilon = \pi - \epsilon /t$. In the case of curved background the growing part of perturbation cannot be interpreted in this way; this fact leads to the conclusion that~\eqref{Unstable_Solution} is a repulsor rather than attractor.

The true solution in evolving background with $P \ll 1$ behaves as follows. At early times the terms with the Hubble parameter are negligible in eq.~\eqref{G_EOM}, and the solution is the Minkowski attractor,

\be
e^{\pi} = \frac{1}{H_* (t_1 - t)},~~~t \to -\infty .
\label{pi_Minkowski}
\ee
Later on, the effect of the Hubble parameter accumulates, and the solution deviates from eq.~\eqref{pi_Minkowski}. However, after a certain transition period, the Galileon field and its time derivatives become so large that the Hubble terms in eq.~\eqref{G_EOM} become negligible again. The solution asymptotes to
\be
e^{\pi} = \frac{1}{H_* (t_* - t)}
\label{pi_flat}
\ee
with some new parameter $t_*$. The latter regime occurs when
\[
|t_* - t| \ll \frac{|t|}{P}.
\]
An example of numerical solution is shown in Fig.1.
\vspace{1 cm}
\begin{center}
\includegraphics[scale=0.9]{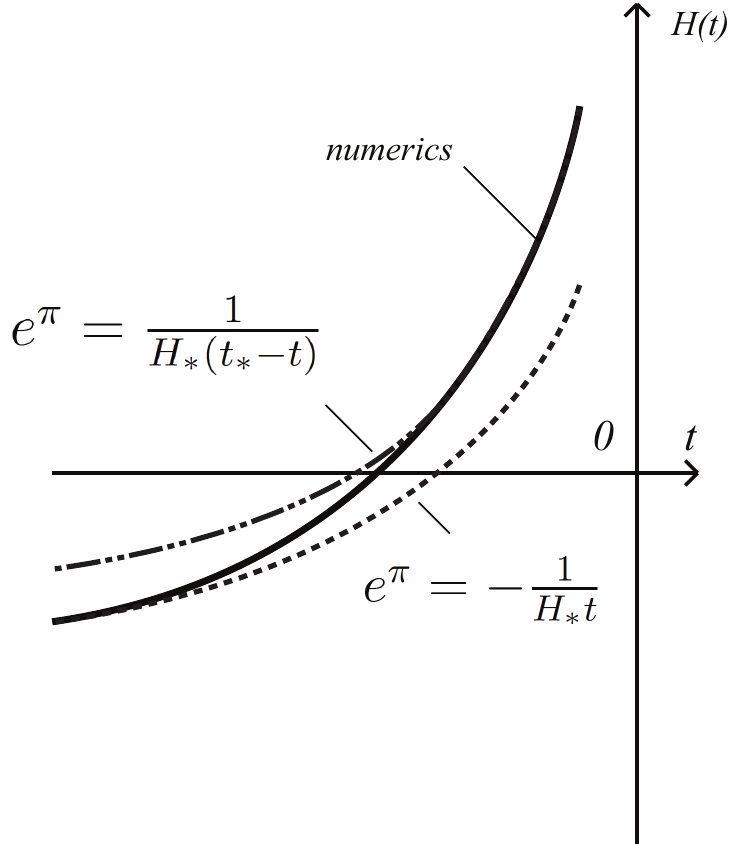}

Figure 1.~\itshape{The Galileon in the background of scalar field. Numerical solution (solid line) and asymptotics $e^{\pi} = -\frac{1}{H_* t}$ (dashed line) and $e^{\pi} = \frac{1}{H_* (t_* - t)}$ (dot-dashed line).}
\end{center}

The solution~\eqref{pi_flat} is valid modulo small corrections, provided that the terms with the Hubble parameter are small in the Galileon field equation. As we will now see, this situation can occur even at the bounce epoch. We call this regime "slow bounce".

\subsection{Slow bounce}
\label{slow_bounce}

The Galileon equation has rather complicated form and the full system of equations cannot be solved exactly. However, there exists a useful approximation. We have seen in Section~\ref{scalar_bckgr} that at times of interest the Hubble terms in the Galileon equation are negligible, and the Galileon rolls just like in Minkowski space-time, see eq.~\eqref{pi_flat}. Likewise, for $M_p c \gg 1$, the scalar field~\eqref{scalar_solution_curved} also exhibits Minkowskian behavior. This suggests that we can make use of perturbative approach in the small parameter $M_p^{-2}$. 

To the leading order in $M_p^{-2}$, the Galileon and scalar field $\phi$ decouple. We use their Minkowski space solutions, eq.~\eqref{pi_flat} and
\be
\phi = - \frac{1}{c} \ln {\frac{2}{c^2 V_0 t^2}},
\label{phi_flat_solution}
\ee
and find that energy densities of the Galileon and scalar field vanish, while their pressures do not:
\be
\rho_G^{(0)} = 0,~~~\rho_{\phi}^{(0)} = 0 \; ,
\label{rho=0}
\ee

\[
p_G^{(0)} = -3 \left( \frac{f}{\Lambda} \right )^3 \cdot (t_* - t)^{-4} \; ,
\]

\[
p_{\phi}^{(0)} = \frac{4}{c^2 t^2} \; .
\]
We then use eq.~\eqref{Raichaudkhuri} to calculate the first-order Hubble parameter:

\be
H^{(1)} = \frac{1}{M_p^2} \left( \frac{2}{c^2 t} + \frac{f^2}{3 H_*^2}\frac{1}{(t_* - t)^3} \right)  \; . 
\label{H_slowbounce}
\ee
Here the integration constant is set equal to zero because of the initial condition $H \to 0$ as $t \to -\infty$. 

Note that $H^{(1)2} \propto M_p^{-4}$, and hence, in view of the Friedmann equation, $\rho_G + \rho_{\phi} \propto M_p^{-2}$. This is consistent with eq.~\eqref{rho=0}: energy densities become non-vanishing at the first order of the expansion in $M_p^{-2}$.

As it stands, eq.~\eqref{H_slowbounce} describes ekpyrotic contraction and bounce. We have to check, however, that bounce occurs within the range of validity of our perturbative treatment.

\subsubsection{Conditions of validity}
\label{Area}

Our perturbative approach is valid provided that the gravitational corrections to the field equations are small. We are interested in the regime when $H$ crosses zero; that happens at time $t_b$ such that
\be
\frac{2}{c^2 t_b} + \frac{f^2}{3 H_*^2}\frac{1}{(t_* - t_b)^3} = 0.
\label{H=0}
\ee
%Let us first consider the equation for the scalar $\phi$:
%\be
%\ddot\phi + c V_0 e^{-c\phi} + 3 H \dot\phi = 0
%\label{Scalar_EOM}
%\ee
It is straightforward to see that the gravitational correction to the field equation for $\phi$ is small provided that
\be
c^2 M_p^2 \gg  1.
\label{first_ie}
\ee
The terms in the Galileon equation~\eqref{G_EOM} that do not include gravity are of the order $\left[ H_*^2 (t_* - t_b)^4 \right]^{-1}$ at $t = t_b$, whereas gravity corrections make two combinations: 
\[
\left[ H_*^2 c^2 t_b (t_b - t_*)^3 \right]^{-1},~~~\left[ H_*^2 c^2 t_b^2 (t_b - t_*)^2 \right]^{-1}. 
\] 
In view of eq.~\eqref{first_ie} they are small, provided that
\be
c M_p |t_b| \gg |t_* - t_b|.
\label{second_ie}
\ee

Let us now study solutions to eq.~\eqref{H=0} to see whether eq.~\eqref{second_ie} is satisfied. Note that our solution has two singular points, $t = t_*$ and $t = 0$, so all physics occurs at $t < 0,~~t < t_*$. First, one can check that for $t_* < 0$, eq.~\eqref{H=0} has one solution in the $t<0$ region, and this solution satisfies the condition~\eqref{second_ie}, so the bounce inevitably occurs. For $t_* > 0$, there are two possibilities:

\textbf{1.} $t_* \gg c f / H_*$. In this case eq.~\eqref{H=0} does not have negative solutions. This means that there is no bounce within the range of validity of our perturbative treatment.

\textbf{2.} $t_* \ll c f / H_*$. This inequality implies that there exist two negative roots of eq.~\eqref{H=0}. One of them is always within the region of validity and corresponds to bounce, whereas the second one describes "rebounce", after which the Universe contracts again. The rebounce occurs at
\be
t_{r} \sim - \frac{H_*^2}{c^2 f^2} t_* ^3.
\nonumber
\label{rebounce_root}
\ee
This second root satisfies eq.~\eqref{second_ie} provided that
\be
t_*^2 > \frac{f^2}{H_*^2 M_p^2}.
\label{rebounce_root_out}
\ee
If the relation~\eqref{rebounce_root_out} is violated, there remains one root of eq.~\eqref{H=0} within the region of validity, and simple bounce occurs.

Three possible types of behavior we have found within our perturbative approach are shown in  Fig.2.
\vspace{1 cm}
\begin{center}
\includegraphics[scale=0.6]{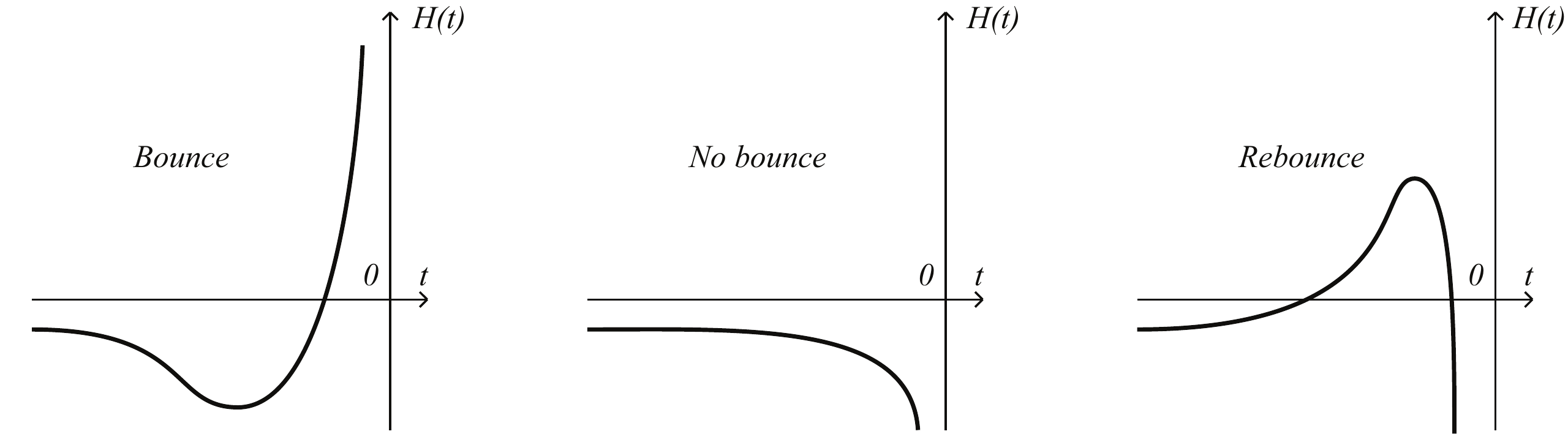}

Figure 2.~\itshape{Three types of behavior in perturbative regime}
\end{center}

\subsection{Numerical results}
\label{numerics}

We have studied numerically the full system of equations describing our model. In the first place, we have checked that numerical solutions coincide with the analytical results of Section~\ref{slow_bounce} within the range of validity of the latter. In particular, once simple bounce (no rebounce) occurs when the perturbative treatment is applicable, the Universe continues to expand in the non-perturbative regime as well. This is illustrated in Fig.~3.

\vspace{1 cm}
\begin{center}
\includegraphics[scale=1.0]{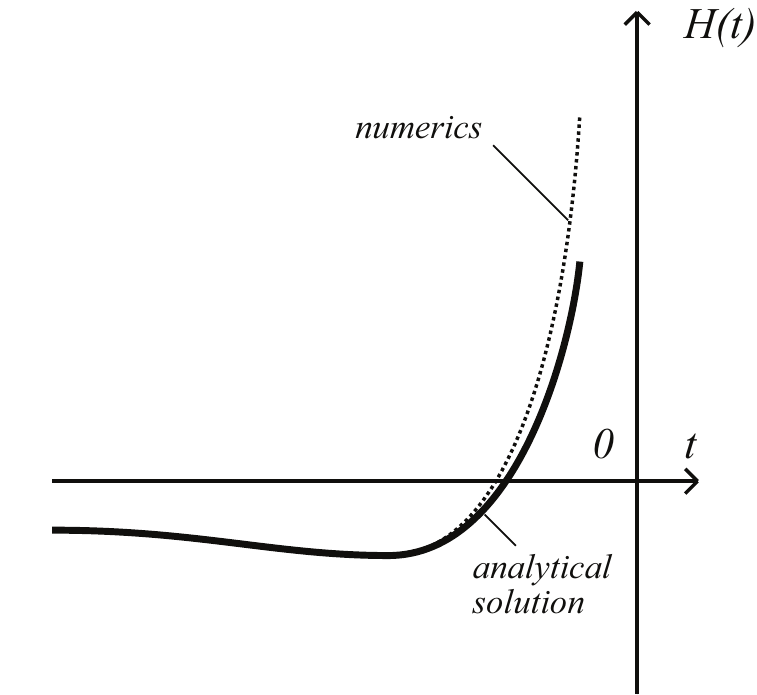}

Figure 3.~\itshape{Analytical solution (solid) vs numerics (dotted)}
\end{center}

However, numerics does not show two other types of behavior: "no bounce" and "rebounce" with subsequent collapse. The bounce is always there. To see what is going on, let us choose some value of $t_* >0$ in eq.~\eqref{pi_flat} treated as the initial condition, and vary the parameter $f$. For small $f$ (case 1 of Section~\ref{slow_bounce}) there is no bounce analytically, but the bounce does occur outside the range of validity of the perturbative treatment (Fig.~4a). For larger $f$ (case 2 of Section~\ref{slow_bounce} with eq.~\eqref{rebounce_root_out} satisfied), there is indeed a rebounce, but after the second contraction stage the Universe bounces again, now in the non-perturbative regime (Fig.~4b). For even larger $f$, when eq.~\eqref{rebounce_root_out} is violated, the rebounce does not occur and the Universe continues to expand after bounce (Figs.~4c,~4d).

\vspace{1 cm}
\begin{center}
\includegraphics[scale=0.7]{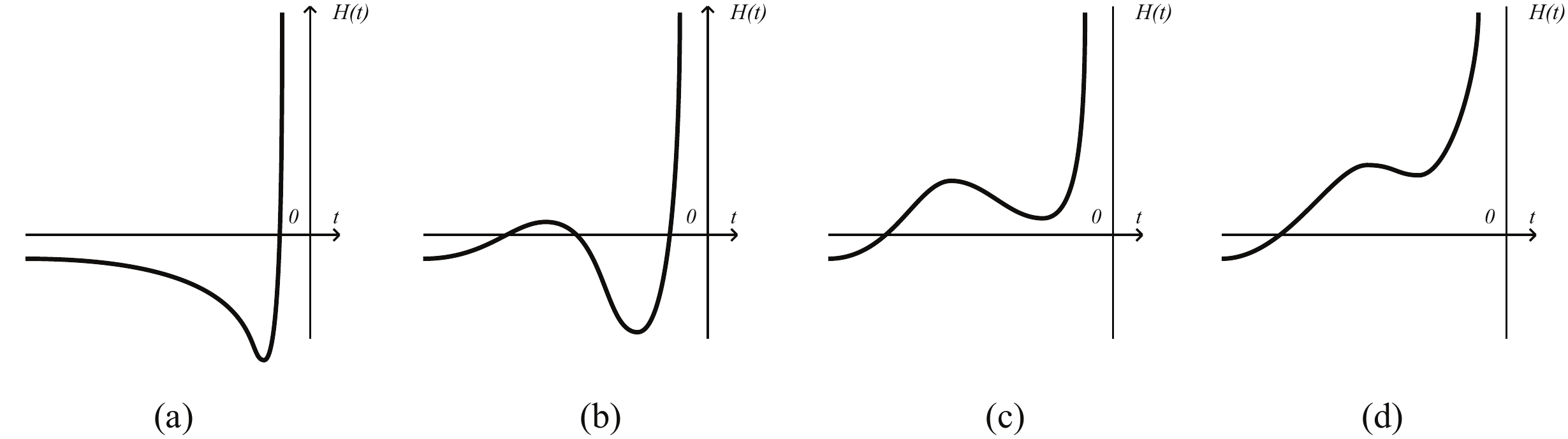}

Figure 4.~\itshape{Numerical results. Behavior of the Hubble parameter depends on the Galileon strength that is determined by the parameter $f$}
\end{center}

At qualitative level, the fact that the Hubble parameter does not decrease indefinitely but starts to grow instead, can be understood in the following way. Suppose that $H(t)$ decreases and has large negative value, so that it dominates in the Galileon equation~\eqref{G_EOM}. One observes that to the leading order in $H$, the Galileon equation reduces to
\be
\ddot \pi = -\frac{3}{2}\dot\pi H.
\label{pi_second_derivative}
\ee
In this approximation, the energy density and pressure of the Galileon field are
\[
\rho_{\pi} = 4 \frac{f^2}{H_*^2} \dot \pi^3 H,~~~p_{\pi} = 2 \frac{f^2}{H_*^2} \dot \pi^3 H.
\]
We plug these expressions into eq.~\eqref{Raichaudkhuri} together with $(p+\rho)_{\phi} = \dot\phi^2$ and obtain
\[
\dot H = -4 \pi G \left( 6 \frac{f^2}{H_*^2} \dot \pi^3 H + \dot\phi^2 \right).
\]
Since the Hubble parameter has large negative value, the system eventually enters the regime in which
\be
\dot\phi^2 \ll |6 \frac{f^2}{H_*^2} \dot \pi^3 H|.
\label{LargeNegativeH_inequality}
\ee
In this case the derivative of the Hubble parameter becomes positive and $H$ begins to grow.

To substantiate this qualitative argument, let us check that the regime~\eqref{LargeNegativeH_inequality} is self-consistent. To this end we take the derivative of that inequality:
\be
2 \dot\phi \ddot\phi \ll |\frac{f^2}{H_*^2} (3 \dot\pi^2 \ddot\pi H + \dot\pi^3 \dot H)|.
\label{LargeNegativeH_inequality_derivative}
\ee
Then we use the field equation for $\phi$ together with eq.~\eqref{pi_second_derivative}, and find that eq.~\eqref{LargeNegativeH_inequality_derivative} transforms back into eq.~\eqref{LargeNegativeH_inequality}. We see that eq.~\eqref{LargeNegativeH_inequality} is consistent with field equations: its left hand side grows slower than the right hand side. Hence, once our system has entered the regime~\eqref{LargeNegativeH_inequality}, it does not leave it.

To end up this section, we note that in a wide range of parameters and initial data, all regimes we have found take place at sub-Planckian energy densities and pressures. However, pushing the parameters and/or initial data, one does find situations when the bounce does not occur before the system  reaches the Planckian parameters. This happens, in particular, for very large positive values of the parameter $t_*$ characterizing the initial data for the Galileon: at the level of classical field equations, the bounce does occur in this case but it cannot be trusted, since pressure exceeds $M_p^4$ before the bounce.

\section{Extended model}
\label{Exit}

In the $\phi-\pi$ system, the Galileon dominates at late times. To be more realistic, one has to assume that the Galileon eventually decays, i.e., its energy density gets transformed into heat, see Ref.~\cite{Levasseur}. We also have to improve on the dynamics of $\phi$, since, as we mentioned above, the potential $V(\phi)$ is unbounded from below. We modify this potential in such a way that it obtains a minimum at finite $\phi$. Our analysis in Sec.~\ref{Bounce} will be valid, provided that the potential remains nearly exponential before and at the bounce epoch, so that our modification affects late time dynamics only. 

Let us denote the depth of the modified potential by $V_{min}(\phi) = - W_0$. To make sure that the cosmological constant vanishes at late times, we also introduce a potential for the Galileon, $W(\pi)$, that asymptotes to $W_0$. The picture of Sec.~\ref{Bounce} holds if it is equal to zero at the stage of bounce and earlier, but grows to the value $W_0$ at late times. In fact, we have seen by numerical analysis that the whole picture often does not change even if the Galileon reaches $W_0$ before the bounce.

In the modified model, the field $\phi$ does not indefinetly roll down. There are several possible types of its late-time evolution. The first one is that $\phi$ reaches the minimum of $V(\phi)$ and settles down there after oscillating for some time, while the Galileon climbs its potential and decays later. The second possibility is that due to the rapid growth of the Hubble parameter (which is driven by the rapidly growing Galileon field), the scalar $\phi$ gets frozen before it reaches the minimum of $V(\phi)$. In that case its evolution stops at some value $\phi_s$ and continues only after the decay of the Galileon field. This situation can in principle lead to the standard inflationary stage, but we will show that this does not happen in a wide range of parameters.

Let us discuss the dynamics using a specific example of modified potentials.

\subsection{Example}
\label{Exit_Example}

Let us choose the scalar potential in the following way (left graph in Fig.~5):
\[
V(\phi) = -V_0 e^{-c\phi} \left [ 1 - \alpha e^{-b\phi}   \right ],~~~b > c.
\]
We also introduce the potential of the Galileon field (right graph in Fig.~5):
\[
W(\pi) = \frac{W_0}{2} \left[ 1 + {\rm tanh}(  q( \pi - \pi_{1} )  ) \right].
\]
Here $q \sim 1$, and $\pi_1$ is the parameter that determines the time when the Galileon starts to climb the potential. We choose $W_0$ to be equal to the depth of the modified potential $V(\phi)$, so the cosmological constant vanishes at late times,
$$
W_0 = V_0 \frac{b}{b+c} \left( \frac{c}{\alpha (b+c)} \right )^{\frac{c}{b}}.
$$

\vspace{1 cm}
\begin{center}
\includegraphics[scale=0.7]{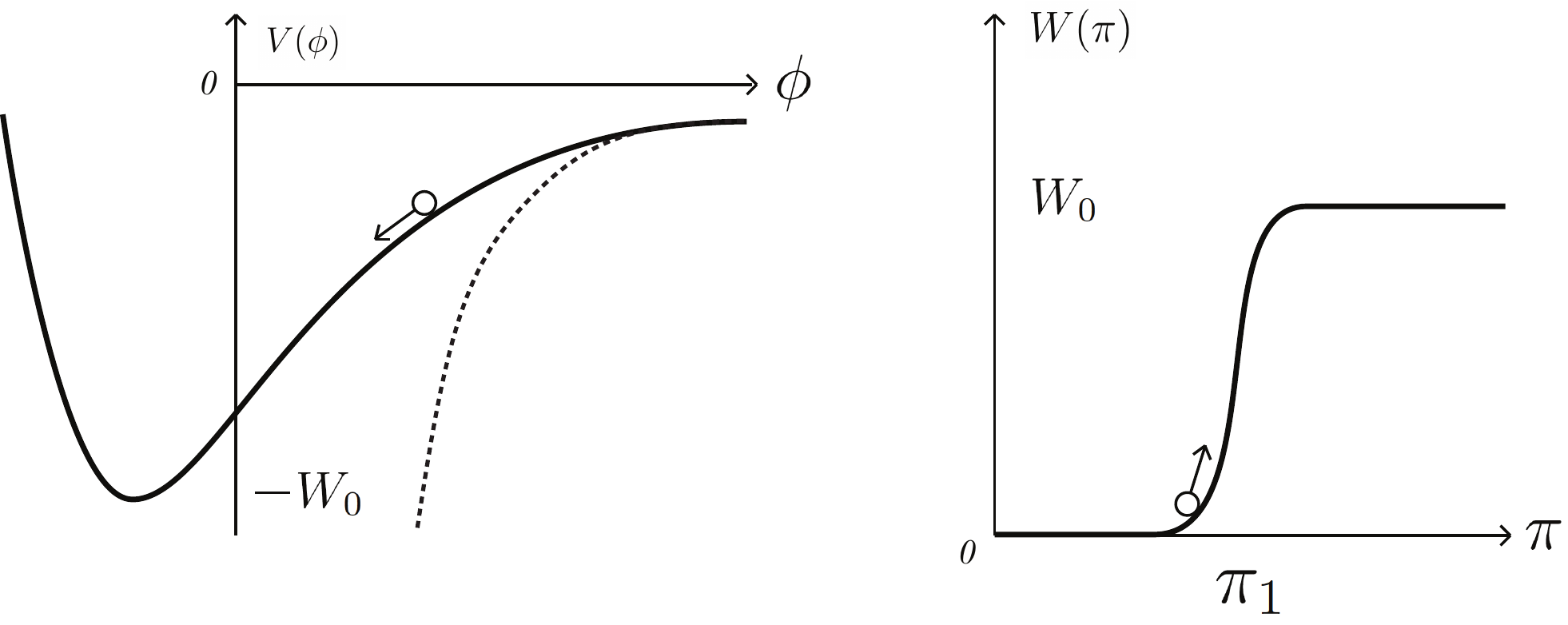}

Figure 5.~\itshape{Modified potentials and the original potential $V(\phi)$ (dotted line).}
\end{center}

Numerical analysis shows that depending on the parameters, the scalar field $\phi$ can either reach the minimum of its potential $V(\phi)$ and oscillate for some time or stop earlier due to the Hubble friction. The first situation does not require further analysis: near the minimum of $V(\phi)$ we are dealing with massive scalar field whose value is well below $M_p$; hence, the model never enters the inflationary regime. We will shortly give an example of the parameter choice yielding this situation, see eq.~\eqref{for_oscillations}.

Less trivial is the case in which the field $\phi$ does not reach the minimum of its potential and becomes frozen at some point $\phi_s = \phi(t_s)$. Let us see that in a wide range of parameters the inflationary slow roll conditions for $V(\phi)$ are violated, so the inflation stage does not occur after the decay of the Galileon. To this end, we consider initial conditions such that $ t_* \gg f/M_p H_*$ for the Galileon, and obtain an estimate for $t_s$. The scalar field equation is
$$
\ddot\phi + c V_0 e^{-c\phi} + 3 H \dot\phi = 0.
$$
The field $\phi$ begins to slow down when
\be
|\ddot\phi| \sim |3 H \dot\phi|.
\nonumber
\label{stop-1}
\ee
Plugging the Minkowski solution~\eqref{phi_flat_solution} for the field $\phi$ into this expression, one gets for freeze-out time 
\be
H \sim \frac{1}{t_s}.
\label{H_at_stop}
\ee
This occurs when the Galileon dominates in the Hubble parameter (we check this below), so that
\be
H \sim \frac{f^2}{3 M_p^2 H_*^2} \frac{1}{(t_* - t_s)^3}.
\label{H_late}
\ee
Combining eqs.~\eqref{H_at_stop} and~\eqref{H_late} one gets
\be
\frac{f^2}{M_p^2 H_*^2} \frac{1}{|t_* - t_s|^3} \sim \frac{1}{|t_s|}.
\label{eq_for_t_s}
\ee
Under our assumption $ t_* \gg f/M_p H_*$, we find that eq.~\eqref{eq_for_t_s} is satisfied when $|t_* - t_s| \ll t_*$, so that
$$
t_* - t_s \sim  \left( \frac{f^2 t_*}{M_p^2 H_*^2} \right )^{\frac{1}{3}}.
$$
Let us check that at this moment the slow bounce regime is still valid for the Galileon field. As we pointed out in Sec.~\ref{Area}, the terms in the Galileon equation that do not include gravity are of the order $\left[ H_*^2 (t_* - t)^4 \right]^{-1}$. Corrections due to the Hubble parameter are $H \cdot \left[ H_*^2 (t_* - t)^3 \right]^{-1}$ and $\dot H \cdot \left[ H_*^2 (t_* - t)^2 \right]^{-1}$. They are small provided that
$$
\frac{1}{(t_* - t_s)^2} \ll \frac{M_p^2 H_*^2}{f^2}.
$$
This inequality is indeed valid for $ t_* \gg f/M_p H_*$.

To check the consistency of our calculation, we observe that the scalar contribution to the Hubble parameter is small at $t\sim t_s$. Indeed, we have
$$
\frac{f^2}{H_*^2 (t_* - t_s)^3} \sim \frac{M_p^2}{t_s} \gg \frac{1}{c^2 t_s}
$$
for $M_p c \gg 1$, so the first term in the right hand side of eq.~\eqref{H_slowbounce} is small compared to the second term.

Now we turn to the main issue of whether the Universe enters the inflationary regime at late times after the Galileon decay. This can happen if the field $\phi$ freezes out at the exponential part of its potential, $\alpha \exp (-b\phi) \ll 1$. This gives

\be
\alpha (c^2V_0 t_s^2)^{-\frac{b}{c}} \ll 1.
\label{V_still_exp}
\ee
In the opposite case, 
\be
t_s \ll \alpha^{\frac{c}{2b}} \frac{1}{c\sqrt{V_0}},
\label{for_oscillations}
\ee
the field $\phi$ stops very near the minimum of its potential, after oscillating about this minimum. As we have already pointed out, this behavior does not result in the inflationary stage.

We proceed with the case~\eqref{V_still_exp}. The inflationary slow roll parameter at $\phi = \phi_s$ is
$$
\eta = M_p^2 \frac{V''}{V + W_0} \sim \frac{(M_p c)^2}{\alpha^{-\frac{c}{b}} e^{c\phi}} = \frac{(M_p c)^2}{\alpha^{-\frac{c}{b}} c^2V_0 t_s^2} .
$$ 
Therefore, the inflationary slow-roll condition is violated provided that
\be
\alpha^{-\frac{c}{b}} c^2 V_0 t_s^2 \ll (M_p c)^2.
\nonumber
\label{no_inflation}
\ee
We recall that $t_s \approx t_*$ and conclude that the late time evolution of our system will not lead to the slow roll inflation at least for the following choice of initial data:
$$
\frac{f}{M_p H_*} \ll t_* \ll \alpha^{\frac{c}{2b}} \frac{M_p}{\sqrt{V_0}}.
$$
Clearly, if the parameters of the action for $\pi$ and $\phi$ are far from Planckian, this range is parametrically wide.

\section{Conclusion}

We have shown that there exists a simple way to construct a bouncing cosmological model which incorporates a contraction epoch with $p \gg \rho$ and bounce due to the Galileon. Notably, all this can happen in the regime where gravity can be treated as perturbation. This is not a necessity, though: we have seen by numerical analysis that the Galileon dynamics is so strong that it always yields the bounce, irrespectively of the value of the Hubble parameter late at the contraction stage. We have made the model more realistic by ensuring that the cosmological constant is negligible after the putative decay of the Galileon and checked that the late-time inflationary epoch is avoided in a wide range of initial data. All this makes models of this type consistent, albeit somewhat baroque, alternative to inflation.

\section{Acknowledgements}
We thank S.~Sibiryakov, M.~Kuznetsov, A.~Vikman, V.~Khlebnikov for helpful discussions. This work has been supported in part by grants RFBR 12-02-00653, RFBR 12-02-31778, NS-5590.2012.2 and grant of Ministry of Science and Education of Russian Federation No. 8412.

\end{document}